\documentclass{article}
\usepackage{enumerate}
\usepackage{multirow}
\usepackage[bottom,flushmargin,hang,multiple]{footmisc}
\usepackage{etoolbox}
\makeatletter
\patchcmd\maketitle{\hb@xt@1.8em}{\hbox}{}{}
\makeatother
\patchcmd{\maketitle}{\@fnsymbol}{\@arabic}{}{}
\usepackage{amsmath}
\usepackage{graphicx}
\usepackage{caption}
\usepackage{subcaption}
\usepackage[margin=1.5in]{geometry}
\usepackage{float}
\usepackage{comment}
\usepackage{pgf}
\usepackage[T1]{fontenc}
\usepackage[sfdefault,scaled=.95,light]{FiraSans}
\usepackage{newtxsf}

\usepackage{url}
\usepackage{apacite}

\begin{document}

\title{An evaluation of machine learning techniques to predict the outcome of children treated for Hodgkin-Lymphoma on the AHOD0031 trial \\
\large A report from the Children's Oncology Group}


\author{C\'edric Beaulac$^a$  \and Jeffrey S. Rosenthal$^b$ \and Qinglin Pei$^c$  \and Debra Friedman$^d$ \and Suzanne Wolden$^e$ \and David Hodgson$^f$ %etc.
}

%\footnote{ University of Toronto,  Department of Statistical Sciences}
%\footnote{ University of Toronto, Department of Radiology Oncology} 
%\author{\makebox[.9\textwidth]{C\'edric Beaulac} \\ University of Toronto  \and Qinglin Pei \\ University of Florida \and Jeffrey S. Rosenthal \\ University of Toronto \and Co-Author3\\Dept  \and David Hodgson \\ University of Toronto }


%\textsc{\institute{C\'edric Beaulac \at
%              University of Toronto \\
%              Tel.: +123-45-678910\\
%              Fax: +123-45-678910\\
%              \email{fauthor@example.com}           %  \\
%           Jeffrey S. Rosenthal \at
%             University of Toronto
%}}

%\institute{    C\'edric Beaulac         \and
 %       Jeffrey S. Rosenthal \at
	%			University of Toronto \\
%				Sidney Smith Hall \\
	%			100 St. George Street, Toronto, Ontario, Canada 
%              Tel.: +123-45-678910\\
%              Fax: +123-45-678910\\
%              \email{fauthor@example.com}           %  \\

\date{\today}

\maketitle

\noindent \textit{$^a$ Department of Statistical Sciences, University of Toronto, Toronto, Canada; $^b$Department of Statistical Sciences, University of Toronto, Toronto, Canada; $^c$Department of Biostatistics, University of Florida, Gainesville, USA; $^d$Department of Pediatrics, Vanderbilt University, Nashville, USA; $^e$Department of Radiation Oncology, Memorial Sloan Kettering Cancer Center, New York, USA; $^f$Department of Radiation Oncology, University of Toronto, Toronto, Canada.}

\begin{abstract}
In this manuscript we analyze a data set containing information on children with Hodgkin Lymphoma (HL) enrolled on a clinical trial. Treatments received and survival status were collected together with other covariates such as demographics and clinical measurements. Our main task is to explore the potential of machine learning (ML) algorithms in a survival analysis context in order to improve over the Cox Proportional Hazard (CoxPH) model. We discuss the weaknesses of the CoxPH model we would like to improve upon and then we introduce multiple algorithms, from well-established ones to state-of-the-art models, that solve these issues. We then compare every model according to the concordance index and the brier score. Finally, we produce a series of recommendations, based on our experience, for practitioners that would like to benefit from the recent advances in artificial intelligence.

\bigskip

%\noindent

\textbf{Keywords} : machine learning, case study, survival analysis, Cox proportional hazard, survival trees, neural networks, variational auto-encoders
\end{abstract}

\pagebreak

\section{Introduction}

There is increasing effort in medical research to applying ML algorithms to improve treatment decisions and predict patient outcomes. In this article, we want to explore the potential of ML algorithms to predict the outcome of children treated for Hodgkin Lymphoma. As we want to minimize the side effects of intensive chemotherapy or radiation therapy, a major clinical concern is how, for a given patient, we can select a treatment that eradicates the disease while keeping the intensity of the treatment, and the associated side effects, to a minimum.

\bigskip

In this article we will introduce multiple ML algorithms adapted to our needs and compare them with the Cox proportional hazard model. As it is the case with many data set within this field, the response variable, time until death or relapse, was right-censored for patients without
events and the data set is of relatively small size (n=1712). From a ML perspective, this can be challenging. The response variable is right-censored for multiple observations but many ML techniques are not designed to deal with censored observations and thus it restricts the techniques we can include in our case study. Another challenged previously mentioned is that medical data sets are usually smaller than those used in ML applications and thus we will have to carefully select algorithms that could perform well in this context.

\bigskip

We will introduce the data set in section 2. In section 3 we will introduce the algorithms tested. Then, in section 4 we will present our experimental set up and our results. Finally, in section 5, we will discuss thoroughly the results, recommend further improvements and introduce open questions.

\section{Data set}
\label{Data}

We have a data set of 1,712 patients, treated on the Children's Oncology Group trial AHOD0031, the largest randomized trial of pediatric HL ever conducted. Each observation represents a patient suffering from Hodgkin Lymphoma. For every patient, characteristics and symptoms have been collected as well as the treatment, for a total of 21 predictors. A table containing information on the predictors is in the appendix. The response is a time-to-event variable registered in number of days. We consider events to be either death or relapse. For patients without events, the response variable was right-censored at time of last seen, which is a well-known data structure in survival analysis. This data set and the data collecting technique are presented in detail by Friedman \& al. \citeyear{Friedman14} who previously analyzed the same data set for other purposes.

\section{Survival Analysis models}
\label{Models}

\subsection{Benchmark : Cox Proportional Hazard Model}

The Cox Proportional Hazard (CoxPH) model \cite{Cox72} serves as our benchmark model. It is widely used in medical sciences since it is robust, easy to use and produce highly interpretable results. It is a semi-parametric model that fits the hazard function, which represents the instantaneous rate of occurrence for the event of interest, using a partial likelihood function \cite{Cox75}.

\bigskip

The CoxPH model fits the hazard function which contains two parts, a baseline hazard function of the time and a feature component which is a linear function of the predictors. The proportional hazard assumption assumes the time component and the feature component of the hazard function are proportional. In other words, the effect of the features is fixed through time. In the CoxPH model, the baseline hazard, which contains the time component, is usually unspecified so we cannot use the model directly to compute the hazard or to predict the survival function for a given set of covariates. 

\bigskip

The main goal of this analysis is to test whether or not new ML models can outperform the CoxPH model. As ML models have shown great potential in many data analysis applications, it is important to test their potential to improve outcome prediction for cancer patients. We would like our selected models to improve upon at least one of the three following problems that are intrinsic to the CoxPH model. Problem (1): the proportional hazard assumption; we would like models that allow for feature effects to vary through time. Problem (2): the unspecified baseline hazard function; we would like models able to predict the survival function itself. Problem (3): the linear combination of features; we would like to use models that are able to grasp high order of interaction between the variable or non-linear combinations of the features.

\subsection{Conventional statistical learning models}

\subsubsection{Regression models}

The first model to be tested is a member of the CoxPH family. One way to capture interactions between predictors in linear models, and thus improve towards problem (3), is to include interaction terms. Since typical medical data sets contains few observations and many predictors, including all interactions usually leads to model saturation.

\bigskip

To deal with this issue we will use a variable selection model. Cox-Net \cite{Simon11} is an extension of the now well-know lasso regression \cite{Hastie09} implemented in the glmnet package \cite{Friedman10} and is the first model we will experiment with. The Cox-Net is a lasso regression-style model that shrinks some model coefficients to zero and thus insures the model is not saturated. The resulting model is as interpretable as the benchmark CoxPH model, but Cox-Net allows us to include all interactions in the base model without losing too many degrees of freedom.

\bigskip

Another approach based on regression models is the Multi-Task Logistic Regression (MTLR). Yu et al. \cite{Chun11} proposed the MTLR model which quickly became a benchmark in the ML community for survival analysis and was cited by many authors \cite{Luck17,Fotso18,Zhao19,Jing19}. The proposed technique directly models the survival function by combining multiple local logistic regression models and considers the dependency of these models. By modelling the survival distribution with a sequence of dependent logistic regression, this model captures time-varying effects of features and thus the proportional hazard assumption is not needed. The model also grants the ability to predict survival time for individual patients. This model solves both problem (1) and (2). For our case study, we used the MTLR R-package \cite{Haider19} recently implemented by Haider.

\subsubsection{Survival tree models}

Decision trees \cite{Breiman84} and random forests \cite{Breiman96,Breiman01} are known for their ability to detect and naturally incorporate high degrees of interactions among the predictors which is helpful towards problem (3). This family of models is well-established and make very few assumptions about the data set, making it a natural choice for our case study.

\bigskip

Multiple adaptations of decision trees were suggested for survival analysis and are commonly referred as survival trees. The idea suggested by many authors is to modify the splitting criteria of decision trees to accommodate for right-censored data. Based on previously published reviews of survival trees \cite{Leblanc95,Bou11}, we have selected four techniques for the case study.  

\bigskip

One of the oldest survival tree models that was implemented in R \cite{R} is the Relative Risk Survival Tree \cite{LeBlanc92}. This survival tree algorithm uses most of the architecture established by CART \cite{Breiman84} but also borrows ideas from the CoxPH model. The model suggested by LeBlanc et al. assumes proportional hazards and partitions the data to maximize the difference in relative risk between regions. This technique was implemented in the rpart R-package \cite{Therneau17}. 

\bigskip

We also selected a few ensemble methods. To begin, Hothorn et al. \citeyear{Hothorn04} proposed a new technique to aggregate survival decision trees that can produce conditional survival function, which solves problem (2). To predict the survival probabilities of a new observation, they use an ensemble of survival trees \cite{LeBlanc92} to determine a set of observations similar to the one in need of a prediction. They then use this set of observations to generate the Kaplan-Meier estimates for the new one. Their proposed technique is available in the ipred R-package \cite{Peters19}. A year later, Hothorn et al. \citeyear{Hothorn05,Strobl07} proposed a new ensemble technique able to produce log-survival time estimates instead. We will test this technique that is implemented in party R-package \cite{Hothorn12,Hothorn19}.

\bigskip

Finally, the latest development in random forests for survival analysis is Random Survival Forests \cite{ishwaran08}. This implementation of a random survival forest was shown to be consistent \cite{Ishwaran10a} and it comes with high-dimensional variable selection tools \cite{Ishwaran10b}. This model was implemented in the randomForestSRC R-package \cite{Ishwaran10}.

\subsection{State-of-the-art models}

\subsubsection{Deep learning models}

The first state-of-the-art model we will experiment with is built upon the most popular architecture of models in recent years: deep neural networks.  Yu et al. \citeyear{Chun11} MTLR model inspired many modifications \cite{Luck17,Fotso18,Zhao19,Jing19} in order to include a deep-learning component to the model. The main purpose is to allow for interactions and non-linear effect of the predictors. For example, Fotso \citeyear{Fotso18,Fotso19} suggested an extension of the MTLR where a deep neural networks parameterization replaces the linear parameterization and Luck et al. \citeyear{Luck17} proposed a neural network model that produces two outputs: one is the risk and one is the probability of observing an event in a given time bin. Unfortunately, the authors for most of these techniques \cite{Luck17,Zhao19,Jing19} did not provide either their code or a package which causes great reproducibility problems and leads to a serious accessibility issue for practitioners. The DeepSurv architecture \cite{Katzman18} proposed by Katzman et. al is a direct extension to the CoxPH model where the linear function of the covariance is replaced by a deep neural network. This allows the model to grasp high-order of interactions between predictors therefore solving problem (3). By allowing for interaction between covariates and the treatment the proposed model provides a treatment recommendation procedure. Finally, the authors provided a Python library available on the first author's GitHub \cite{Katzman17}.

\subsubsection{Latent-variable models}

The final model is a latent-variable model based on the  Variational Auto-Encoder (VAE) \cite{Kingma13,Kingma17} architecture. Louizos et al. \citeyear{Louizos17} recently suggested a latent variable model for causal inference. The latent variables allow for a more flexible observed variable distribution and intuitively model the hidden patient status. Inspired by this model and by the recommendation of Nazbal et al. \citeyear{Nazbal18} we implemented a latent variable model \cite{Beaulac18} that adapts the VAE architecture for the purposed of survival analysis. This Survival Analysis Variational Auto-Encoder (SAVAE) uses the latent space to represent the patient true sickness status and can produce individual patient survival function based on their respective covariates which should solve problem (1), (2) and (3).

\section{Data analysis}

\subsection{Evaluation metrics}

We will use two different metrics to evaluate the various algorithms, both are well established and they evaluate different properties of the models. First, the concordance index \cite{Harrell96} is a metric of accuracy for the ordering of the predicted survival time or hazard. Second, the brier score \cite{Graf99} is a metric similar to the mean squared error but adapted for right-censored observations.

\subsubsection{Concordance Index}

The concordance index (\textit{c}-index) was proposed by Harrell et al. \citeyear{Harrell96}. It is one of the most popular performance measures for survival problems \cite{Steck07,Chen12,Katzman17} as it elegantly accounts for the censored data. It is defined as the proportion of all usable patient pairs in which the predictions and outcomes are concordant. Pairs are said to be concordant if the predicted event times have a concordant ordering with the observed event times.

\bigskip

\begin{figure}
\begin{center}
\includegraphics[width=5cm,height=5cm]{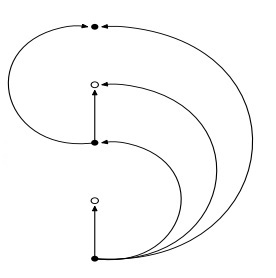}
\caption{Steck et al.(2008) graphical representation of the c-index computation. Filled circle represents observed points and empty circle represents censored points. This figure illustrates the pairs of points for which an order of events can be established.\label{cin}}
\end{center}
\end{figure}

Recently Steck et al. used the \textit{c}-index directly as part of the optimization procedure \cite{Steck07}, their paper also elegantly presents the \textit{c}-index itself using graphical models as illustrated in figure \ref{cin}. In their article it is defined as the fraction of all pairs of subjects whose predicted survival times are correctly ordered among all subjects that can actually be ordered. We expect a random classification algorithm to achieves a $c$-index of 0.5. The further from 0.5 the $c$-index is the more concordant pairs of predictions the model has produced. A $c$-index of 1 indicates perfect predicted order.

Figure \ref{cin} illustrates when we can compute the concordance for a pair of data points; this is represented by an arrow. We can evaluate the order of events if both events are observed. If one of the data points is censored, then concordance can be evaluated if the censoring for the censored point happens after the event for the observed point. If the reverse happens, if both points are censored or if both events happen exactly at the same time then we cannot evaluate the concordance for that pair.

\subsubsection{Brier Score}

The Brier score established by Graf et al. \citeyear{Graf99} is a performance metric inspired by the mean squared errors (MSE). For a survival model it is reasonable to try to predict $P(T>t|X=x)= S(t|X=x)$ the survival probabilities a time $t$ for a patient with predictors $x$. In Graf's notation, $\hat\pi(t|x)$ is the predicted probability of survival at time $t$ for a patient with characteristics $x$. These probabilities are used as predictions of the observed event $y = \mathbf{1}(T > t)$. If the data contains no censoring, the simplest definition of the Brier Score would be

\begin{equation}
\text{BS}(t) = \frac{1}{n} \sum_{i=1}^n (\mathbf{1}(T_i >t) - \hat\pi(t|x_i))^2.
\end{equation}

Assuming we have a censoring survival distribution $G(t) = P(C > t)$ and an associated Kaplan-Meier estimated $\hat G(t)$. For a given fixed time $t$ we are facing three different scenarios :

\smallskip

Case 1: $T_i > t$ and $\delta_i =1$ or $\delta_i =0$

\smallskip

Case 2: $T_i < t$ and $\delta_i =1$

\smallskip

Case 3: $T_i < t$ and $\delta_i =0$,

\noindent where $\delta_1 = 1$ if the event is observed and $0$ if it is censored. For case 1, the event status is 1 since  the patient is known to be alive at time $t$; the resulting contribution to the Brier score is $(1 - \hat\pi(t|x_i))^2$. For case 2, the event occurred before $t$ and the event status is equal to $\mathbf{1}(T_i > t) = 0$ and thus the contribution is $(0 - \hat\pi(t|x_i))^2$. Finally, for case 3 the censoring occurred before $t$ and thus the contribution to the Brier score cannot be calculated. To compensate for the loss of information due to censoring, the individual contributions have to be reweighed in a  similar way as in the calculation of the Kaplan-Meier estimator leading to the following Brier Score 
\begin{equation}
\text{BS}^c(t) = \frac{1}{n} \sum_{i=1}^n \left((0 - \hat\pi(t|x_i))^2 \mathbf{1}(T_i < t,\delta_i =1)(1/\hat G(T_i)) + (1 - \hat\pi(t|x_i))^2\mathbf{1}(T_i > t)(1/\hat G(t))\right).
\end{equation}

\subsection{Comparative results}

The data set introduced in section \ref{Data} was imported in both R \cite{R} and Python \cite{Python}. To evaluate the algorithms we randomly divided the data set into 1500 training observations and 212 testing observations. The models were fit using the training observations and the evaluation metrics were computed on the testing observations.

\bigskip

As mentioned in the previous sections, the CoxPH benchmark and the conventional statistical learning models were all tested in the R language \cite{R}. They were relatively easy to use with very little adjustment needed and clear and concise documentation. The computational speed of these algorithms was fast enough on a single CPU so that we could perform 50 trials. The state-of-the-art techniques needed a deeper understanding of the model as they contain many hyper-parameters that require calibration. They were also slower to run on a single CPU.

\begin{figure}
\centering
\begin{subfigure}{0.95\textwidth}
\includegraphics[width=13cm]{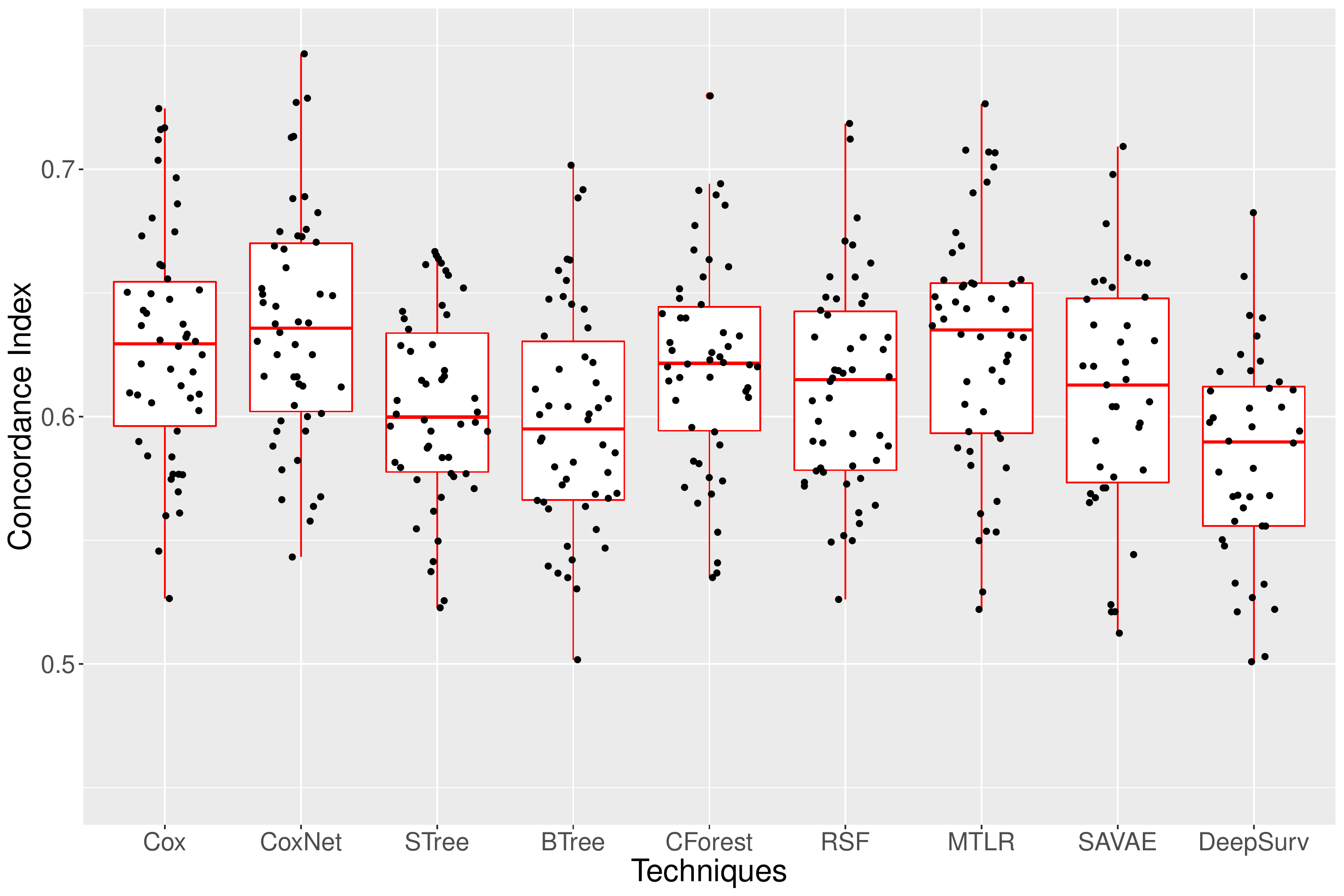}
\caption{Boxplots and Sinaplots of the $c$-index (higher the better).\label{Cindex}}
\end{subfigure}

\begin{subfigure}{0.95\textwidth}
\includegraphics[width=13cm]{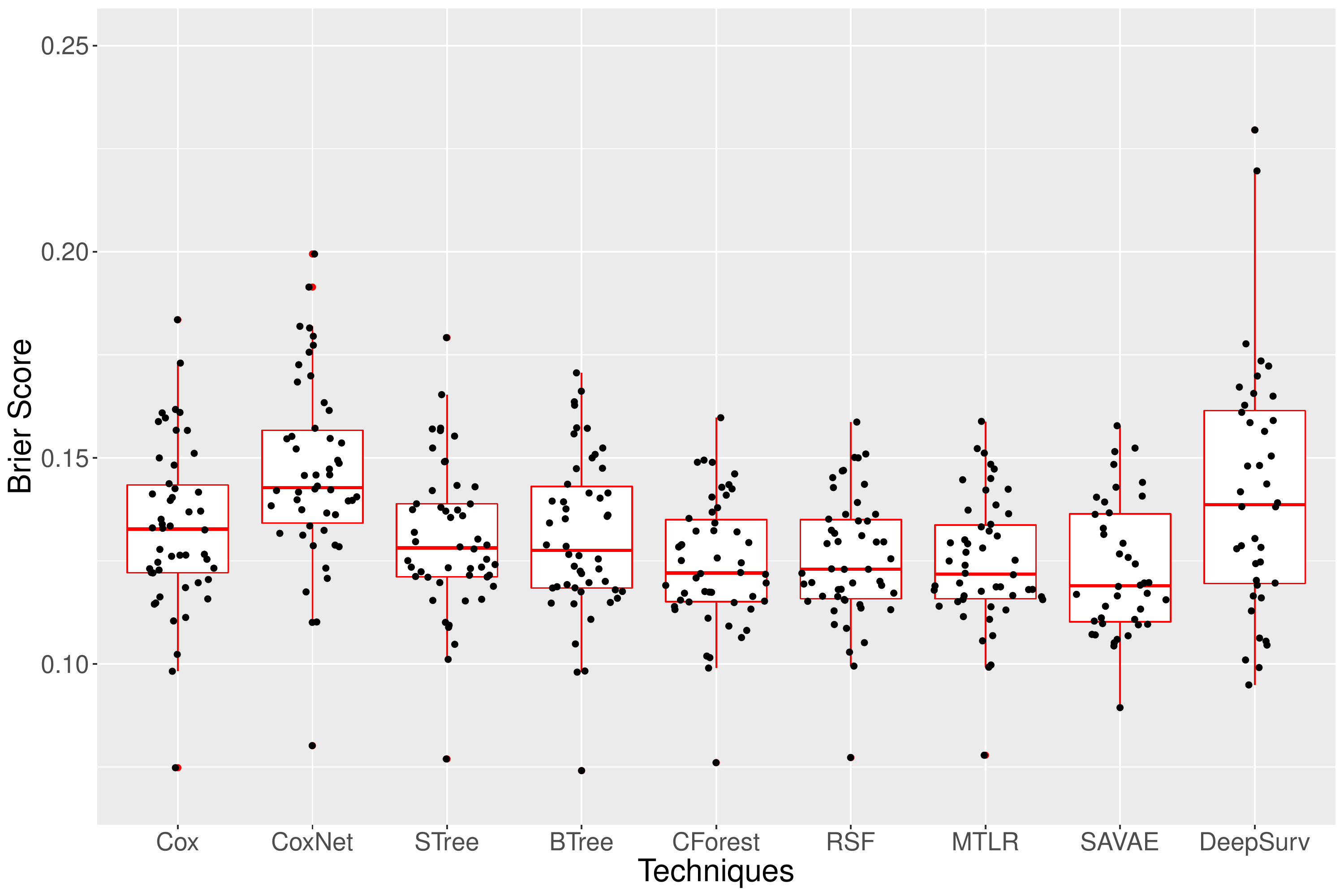}
\caption{Boxplots and Sinaplots of Brier Scores evaluated at 3 years  (lower the better) \label{Brier}}
\end{subfigure}
\caption{Results from our experiments.\label{results_surv}}
\end{figure}

Figure \ref{Cindex} illustrates Sinaplots \cite{Sidiropoulos17}  with associated Boxplots of the $c$-index for the CoxPH model and the 8 competitors. We used standard boxplots on the background since they are common and easy to understand. The sinaplots superposed on them represent the actual observed metric values and convey information about the distribution of the metrics for a given technique. As mentioned earlier c-index ranges from 0.5 to 1 where a $c$-index of 1 indicates perfect predicted order. According to figure 2, it seems no model clearly outperforms another. It seems like MTLR is the best-performing model but the difference is not statistically significant.

\bigskip

Since the Brier score is a metric inspired by the mean squared error, it ranges from 0 to 1 and the lower the Brier score is the better the technique. In figure \ref{Brier} we once again observe that none of the new techniques drastically outperforms any CoxPH. SAVAE has the lowest Brier score and significantly outperforms the CoxPH benchmark.  However, as shows in Figure \ref{Brier} none of the new techniques drastically outperform CoxPH. 

\section{Takeaways and Recommendations}

The previous section demonstrates that the new ML methods offers very little improvement compared to the benchmark CoxPH model according to our two designated performance metrics when patient clinical characteristics that are typically collected in clinical trials are used as predictor variables. This is an important result as we need to evaluate the abilities of ML techniques to solve real-life data problems, and to illuminate the changes in clinical data collection that will have to occur for ML methods to be used to greatest effect in assisting outcome prediction and treatment.

\bigskip

Similar results on real-life data sets are observed in article presenting methodologies \cite{Fotso18,Luck17,Jing19} where the proposed techniques provide non-significant improvements over simple models such as CoxPH. Christodoulou et al. \citeyear{Christodoulou19} recently performed an exhaustive review of 927 articles that discuss the development of diagnostic or prognostic clinical prediction models for binary outcomes based on clinical data. The authors of the review noted the overall poor comparison methodologies and the lack of significant difference between a simple logistic regression and state-of-the-art ML techniques in most of recent years publications. These results are supported by Hand \citeyear{Hand06} who discussed in detail the potential strength of the simple models compared to state-of-the-art ML models. This raises an important question our case study highlights: is it worth using more complex models for a slight improvement?
 
\bigskip

The alternative we proposed in section 3 are all more complicated than CoxPH in various ways. Most of the new techniques require deeper knowledge of the algorithm behaviors to correctly fix the many hyper-parameters. They can produce less interpretable results due to model complexity, and often require more computing power. Indeed, if the CoxPH model can be fit in seconds, most of the conventional statistical learning models take minutes to fit and the state-of-the-art models take hours. Finally, many of the new techniques are not widely accessible or standardized. As an open language, Python offers very little support to users and the libraries are not maintained, not standardized and come with dependency issues.

\bigskip

Hand \citeyear{Hand06} demonstrates the high relative performances of extremely simple methods compared to complex ones and mathematically justifies his argument. He also discusses how these slight improvements over simple models might be undesirable as they might be attributed to overfitting which would cause reproducibility issues on new data sets. These slight improvements might also be artificial as they were achieved only because the inventors of these techniques were able to obtain through much effort the best performance from their own techniques and not the methods described by others. Overall if the improvements over simple techniques are small, perhaps they are simply not an improvement and this argument seems to be supported by both our case study and the recent review of Christodoulou et al. \citeyear{Christodoulou19}. We recommend that practitioners keep their expectations low when it comes to some of these new models. 

\bigskip
 
In contrast, significant improvements for diagnostic tasks have been accomplished using A.I. in recent years \cite{Liu17,Rodriguez19,Rodriguez19-2} and thus we ask ourselves what caused this difference ? There is a major difference in the style of data sets that were available. In the cited articles, images (mammographic, gigapixel pathology image, MRI scans) are analyzed using deep convolutional neural networks (CNN) \cite{Goodfellow16}. Models such as CNN were developed because a special type of data was available and none of the current tools were equipped to analyze it. Conventional techniques such as logistic regression or CoxPH are not able to grasp the signal in images, which contains a large number of highly correlated predictors that individually contain close to no information but analyzed together contain a lot. As a matter of fact, the greatest strength of these models is that they are able to extract a lot of information from a rich, but complicated, data set.

\bigskip

In our case study, the \textit{stratum} predictor was a binary predictor indicating if the patient had a rapid early response to the first rounds of chemotherapy. Computed-tomography (CT) scans of the affected regions were analyzed before and after the first round of treatments and this rich information was transformed into a simple binary variable. This practice is common: even in ongoing trials, patients' characteristics continue to be collected manually (often on paper forms), which dramatically limits the capacity to capture the full range of potentially useful data available for analysis. As new tools are established to extract information from ever growing, both in size and complexity, data sets, clinical trialists have to rethink how they gather data and transform it to make sure that no information is lost in order to utilize these new tools. It seems like extracting and keeping as much information as possible and having a data-centric approach where the model is designed to analyze a specific style of data were some of the factors in the success of CNNs.

\section{Conclusion}

In this article, we have identified a series of statistical and ML techniques that should alleviate some of the flaws of the well-known CoxPH model. These models were tested against a real-life data set and provided little to no improvement according the $c$-index and the Brier score. Although one might anticipate that these techniques would have increased our prediction abilities, instead the CoxPH performed comparably to modern models. These results are supported by other articles with similar findings.

\bigskip

It would be advantageous to try to theoretically understand when the new techniques should work and when they should not. As it currently stands, authors are not incentivized to discuss the weakness of their techniques and it actually slows scientific progress. It is imperative that we try to understand when some of the newest technique perform poorly and shed the light on why it is the case. It is also important to understand what made some of these new techniques successful. For example, it seems that CNNs were successful since the model was specifically built for images, a special type of data that was previously hard to handle but contained a large amount of information.

\clearpage

\section*{Funding details}

Research reported in this work was supported by the Children's Oncology Group; by the National Cancer Institute of the National Institute of Health under the National Clinical Trials Network (NCTN) Operations Center Grant U10CA180886, the NCTN Statistics and Data Center Grant U10CA180899; by St. Baldrick's Foundation; the Natural Sciences and Engineering Research Council of Canada; and by the Ontario Graduate Scholarships.

\bigskip

The content is solely the responsibility of the authors and does not necessarily represent the official views of the Children's Oncology Group, the National Institute of Health, or St. Baldrick's Foundation.

\section*{Disclosure statement}

No potential conflict of interest was reported by the authors.

\clearpage

\section*{Appendix}

\renewcommand\arraystretch{0.7}

\begin{table}[ht]
\begin{center}
\begin{tabular}{ ||p{4.5cm}|p{2.5cm}|p{8cm}||  }
\hline
Variable & Type & Description \\
\hline
\hline
 agedxyrs   & Continuous   & Age of the patient at the start of the treatment \\
 gender &   Binary  & Biological gender  \\
 stage & Categorical & Cancer stage ranging from 1 to 4\\
 b\_symptoms    & Binary & Presence of B symptoms \\
 bulk\_disease &   Binary  & Presence of Bulk disease \\
 extralymphatic\_disease& Binary  & Presence of Extralymphatic disease  \\
 fever & Binary  & Presence of recurrent fever \\
  night\_sweats &   Binary  & Presence of night sweats \\
 weight\_loss & Binary &  Presence of significant weight loss (> 10\%) \\
nodal\_aggregate    & Binary & Presence of a nodal aggregate \\
 mediastinal\_mass &   Binary  & Presence of a mediastinal mass\\
 esron& Continuous  &  Erthroctye sedimentation rate (mm/hr) \\
 istnon & Continuous  & Number of involved nodal sites \\
  histology & Categorical  & Histology (LP,LD,NS,MC, unknown)  \\
 albon & Continuous & Albumin (g/dL)\\
 hgbon    & Continuous & Hemoglobin(g/dL)\\
 amend &   Binary  &\\
 stratum & Binary  & Rapid early response to first treatment   \\
morpho\_icdo & Categorical  & ICD-O Morphology codes \\
 RT & Binary  & Treatment variable: Radiotherapy  \\
DECA & Binary  & Treatment variable: Intensive Chemotherapy  \\
 \hline
\end{tabular}
\caption{Predictor variables and description}
\end{center}
\end{table}

\clearpage

\bibliographystyle{apacite} 
\bibliography{mybibfile}

\end{document}